\documentclass[manuscript]{aastex62}
\usepackage{multirow}
\usepackage{amsmath}
\usepackage{hyperref}

\graphicspath{{./}{figures/}}
\newcommand{\tess}{\emph{TESS}}
\newcommand{\kepler}{\emph{Kepler}}
\newcommand{\jwst}{\emph{JWST}}
\newcommand{\hst}{\emph{HST}}
\newcommand{\plato}{\emph{PLATO}}
\newcommand{\ariel}{\emph{ARIEL}}
\newcommand{\cheops}{\emph{CHEOPS}}

\shorttitle{\tess{} Phase Curves}
\shortauthors{Mayorga, L.~C. et al.}

\begin{document}

\title{Reflected Light Phase Curves in the \tess{} Era}
\author[0000-0002-4321-4581]{L.~C. Mayorga}
\affiliation{Center for Astrophysics \textbar{} Harvard \& Smithsonian, 60 Garden St, Cambridge, MA 02138, USA}
\author{Natasha E. Batalha}
\affiliation{Department of Astronomy \& Astrophysics, University of California, Santa Cruz, Santa Cruz, CA 95064, USA}
\author{Nikole K. Lewis}
\affiliation{Department of Astronomy and Carl Sagan Institute, Cornell University, 122 Sciences Drive, Ithaca, NY 14853, USA}
\author{Mark S. Marley}
\affiliation{NASA Ames Research Center 245-3, Moffett Field, CA 94035, USA}

\begin{abstract}
The reflected light signal from a planet throughout its orbit is a powerful probe of a planet's atmospheric properties. There are a number of planets that are amenable to reflected light phase curve studies with present and future space-based instrumentation and here we assess our ability to characterize these worlds. Using simulated \tess{} populations we identify \replaced{nine}{the Nine, a set of} archetypal exoplanets with the potential to be bright in reflected light, because of their radii and proximity to their star, while still being cool enough to have minimal thermal contamination at optical wavelengths. For each planet we compute albedo spectra for several cloud \added{and atmosphere} assumptions \added{(e.g. thermochemical equilibrium, solar composition)}. We find that in the \tess{} bandpass the estimated contrast at optical wavelengths is typically $<$10~ppm except for the brightest, largest, or closest in planets with the highest lofted clouds where contrast can reach a few tens of ppm. Meanwhile, in a bluer bandpass (0.3--0.5~$\mu$m) the estimated contrast can be as high as 150~ppm but typically 10--50~ppm. In the temperature range of interest, planets with the highest, most extensive cloud decks are generally darker at bluer wavelengths than cloudless planets because of the low single scattering albedos of their primary \added{condensate} constituents. \replaced{We find that}{Our models suggest that} Neptune-sized planets with \added{relatively} low insolation and small semi-major axes are the most conducive to reflected light phase curve studies in \tess{}.
\end{abstract}


\section{Introduction}
The \emph{Transiting Exoplanet Survey Satellite} (\tess{}) mission is expected to find a population of extrasolar planets that \added{encompasses planets that are} generally closer to Earth\added{'s radius and equilibrium temperature} and \replaced{orbits}{in orbit about} cooler stars than those planets found by the \emph{Kepler Space Telescope} and other transiting planet surveys \citep{Borucki2010, Ricker2014}. To accomplish this task, \tess{} is conducting an all sky survey with a redder bandpass than that utilized by \kepler{}. Since the full orbital phase curve of an exoplanet can provide a remarkable number of constraints on the atmospheric structure and dynamics of the planet, it is worthwhile to consider the prospects for such science with \tess{}. In thermal emission the typical quantities measured from a phase curve are the \added{day-/}night-side temperature, and, if any, the offset of the hot spot from the sub-solar point \citep{Parmentier2017}. 

Reflected light phase curves \citep{Seager2000} measure the longitudinal variation of the albedo and the scattering properties of atmospheric aerosols. \added{Reflected light observations are thus very sensitive to the scattering properties (from gas and aerosols and including multiple scattering) of a given planet's atmosphere \citep{Marley1999, Sudarsky2000, Cahoy2010}.} The scattering properties in the atmosphere are \deleted{in turn} controlled by composition of clouds, particle size, pressure or depth of the scattering layer, and more. \added{Consequently in the process of scattering light from its host star, a planet reveals numerous clues about the structure of its atmosphere and key properties, including particle size and vertical distribution, of aerosols that often mute transmission observations. If any TESS discovered planets are amenable to detection in reflected light they will furthermore provide an excellent test bed for validation of models that will later be applied to reflected light direct imaging observations \citep[e.g.,][]{Batalha2019}.}

\kepler{}/K2 measured the albedo of a number of hot Jupiters \citep{Angerhausen2015, Esteves2015, Niraula2018}, and determined that they are typically dark in the \kepler{} bandpass, as expected for cloudless worlds \citep{Marley1999, Sudarsky2000}, with a few notable exceptions like Kepler-7~b, whose inhomogeneous cloud coverage raised its albedo and led to a phase offset \citep{Demory2013, Demory2011}. The theoretical work of \citet{Parmentier2016} showed that the direction of the phase offset is brought about by the temperature of the planet as the planet transitions from thermal emission to reflection dominated in the observational bandpass and further predicts that most hot Jupiters will have cloudy night-sides. 

\added{Spectrophotometric constraints on albedo have also been obtained for some hot Jupiters from other observatories. \citet{Evans2013} measured a relatively bright albedo of HD~189733~b shortward of 450~nm ($A_{\rm g} = 0.4$)  that fell to $A_{\rm g} \leq 0.12$ at longer wavelengths. For its much hotter cousin WASP-12~b only an upper limit of ($A_{\rm g} \leq 0.064$) could be placed \citep{Bell2017}. From the ground \citet{Rodler2010} ruled out an albedo, $A_{\rm g} \geq 0.4$ for $\tau$~Boo~b and there have been measurements of the reflected light from 51~Peg~b suggesting $A_{\rm g} = 0.5$ (see, \citealp{Martins2015, Martins2018,Borra2018}).} 

\added{Many have pointed out the potential for reflected \replaced{observations}{light phase curve characterization} of exoplanets with spaced-based facilities like {\it MOST }{\it CoRoT}, {\it Kepler}, {\it TESS}, {\it CHEOPS}, and {\it PLATO} (e.g., \citealp{Kane2016, Demory2013, Esteves2013, Esteves2015, Munoz2015, Serrano2018}). However the bandpasses from white light photometric missions include a mix of thermal emission and reflected light, an especially problematic situation for hot Jupiters. This can be potentially mitigated through a combination of observing bandpasses (as in \citealp{Placek2016}). There is thus a need to identify which classes of newly discovered planets are most favorable for phase curve followup.}

The \tess{} mission will yield a plethora of planets around bright stars \citep{Ricker2014}. Each \tess{} discovery must be assessed and ranked for subsequent follow-up and characterization with the \emph{James Webb Space Telescope} (\jwst{}) and other observatories. A number of metrics \citep[Zellem et al. in prep]{Kempton2018, Morgan2018} have been generated to determine the threshold for which characterization will be amenable with transmission spectroscopy with \jwst{}. Here, we examine the expected contrast ratios of a potential population of \tess{} planets \citep{Sullivan2015, Barclay2018} in reflected light.

\deleted{As the search for habitable planets continues to move towards smaller and cooler stars, observatories are shifting to longer wavelengths where there are diminished prospects for detecting planets in reflected light. It has been demonstrated that the reflected and thermally emitted light from the atmosphere of a hot Jupiter can be disentangled to more precisely constrain the atmospheric properties through the combination of \tess{} and \kepler{} data \citep{Placek2016}. The all-sky survey \tess{} will complete will result in many new planet discoveries that may require the emitted light to be disentangled from the reflected light and we assess the capabilities of present and future space-based instrumentation, such as \jwst{}/NIRISS-SOSS, \hst{}/WFC3-UVIS, \plato{}, \cheops{}, and \ariel{} to detect and characterize planets in reflected light.}

\replaced{Here we aim t}{T}o consider the prospects for measurement and interpretation of planets in reflected light with \tess{}\replaced{.}{, i}n \autoref{sec:sample}, we outline the determination of the representative sample of planets that are then modeled in \autoref{sec:models}. We discuss the results of the albedo models and the expected contrast in \autoref{sec:results} and conclude in \autoref{sec:conc}.

\section{Sample Selection}
\label{sec:sample}
\citet{Sullivan2015} and \citet{Barclay2018} have created a sample of possible \tess{} planets based on what we know of the instrument, the stars in the galaxy, and planet occurrence rates. Not all of these will be good candidates for reflected light measurements. \added{Giant planets in \tess{} are expected to have a fairly high false positive rates and thus giant planet numbers are the most uncertain (Barclay, Twitter\footnote{https://twitter.com/mrtommyb/status/990259755124953088?s=20}\footnote{https://twitter.com/mrtommyb/status/990258470174814208?s=20}). We use both samples to be agnostic and present a more comprehensive sample of candidates.} Our goal is to determine the classes of planets in the dataset that have the potential to exhibit \added{detectable} reflected light phase curves, how they might group in parameter space, and then determine what kinds of atmospheres they may have and their reflectivity for eventual characterization with follow-up programs.

The studies of \citet{Sullivan2015} and \citet{Barclay2018} provide a number of useful parameters, such as planet size, orbital period, and various stellar properties. From these parameters, we compute additional orbital and planetary properties and finally estimate the reflected light and emitted light signals of all the planets in both samples.

\subsection{The Selected Sample}
To estimate the reflected light and emitted light ratios for the planets, we require the equilibrium temperature of the planet, its orbital semi-major axis, its radius ($T_{\rm eq}, a, R_{\rm p}$), and the stellar radius and temperature ($R_*, T_{\rm eff}$). It is relatively straightforward to compute the equilibrium temperature of the planet given the effective temperature of the star, the radius of the star, the semi-major axis of the planet, and the planetary Bond albedo, $A_B$. The \citet{Barclay2018} sample, at the time of this work, did not list the effective temperature of the star, $T_{\rm eff}$. We computed $T_{\rm eff}$ from the listed $V-Ks$ colors and interpolated onto the grid of \citet{Pecaut2013}.

We assume a typical Bond albedo to be 0.3 \citep{Marley1999} and the geometric albedo spectrum to be flat across all wavelengths. We compute the semi-major axis, $a$, for the \citet{Sullivan2015} sample by assuming circular orbits and from the relation between insolation, $S$, $T_{\rm eff}$, and $R_*$, assuming no internal heat flow,
\begin{align}
S &= \frac{\sigma R_{*}^2 T_{\rm eff}^4}{a^2}\\
a &= R_* T_{\rm eff}^2 \sqrt{\frac{\sigma}{S}},
\end{align}
where $\sigma$ is the Stefan-Boltzmann constant. Thus, \added{under the assumption of complete heat redistribution,} the equilibrium temperature of the planet was computed for all planets in either sample as,
\begin{equation}
T_{\rm eq} = T_{\rm eff} (1-A_{\rm B})^{1/4} \sqrt{\frac{R_*}{2a}}.
\end{equation}

To ensure we do not have significant thermal energy contamination in the \tess{} filter, we have only selected planets where the expected ratio between the reflected light, $F_{\rm R}$, and the thermally emitted light, $F_{\rm E}$, is larger than 10 within the \tess{} bandpass. We also restrict the ratio of the reflected light component to the host star's flux to values large enough to be detectable under reasonable conditions. The \tess{} noise floor on hourly timescales is assumed to be 60~ppm \citet{Ricker2014}, we have imposed a cutoff of $F_{\rm R}/F_*>30$~ppm \added{which is further motivated by the initial error estimates and limits found in \citet{Shporer2019}}.

We compute the reflected light of the predicted planets using the standard equation 
 \begin{equation}
    \frac{F_{\rm R}}{F_*} = A_{\rm g} \left(\frac{R_{\rm p}}{a}\right)^2,
\end{equation}
where $R_P$ is the radius of the planet and we assume a wavelength invariant geometric albedo, $A_g = 0.3$\footnote{In general $A_{\rm B} \ne A_{\rm g}$ because even if $A_{\rm g}$ is constant with $\lambda$ they differ by the phase integral. Here, we simply equate the two in order to make these cuts in parameter space.}.

\begin{figure*}
    \plotone{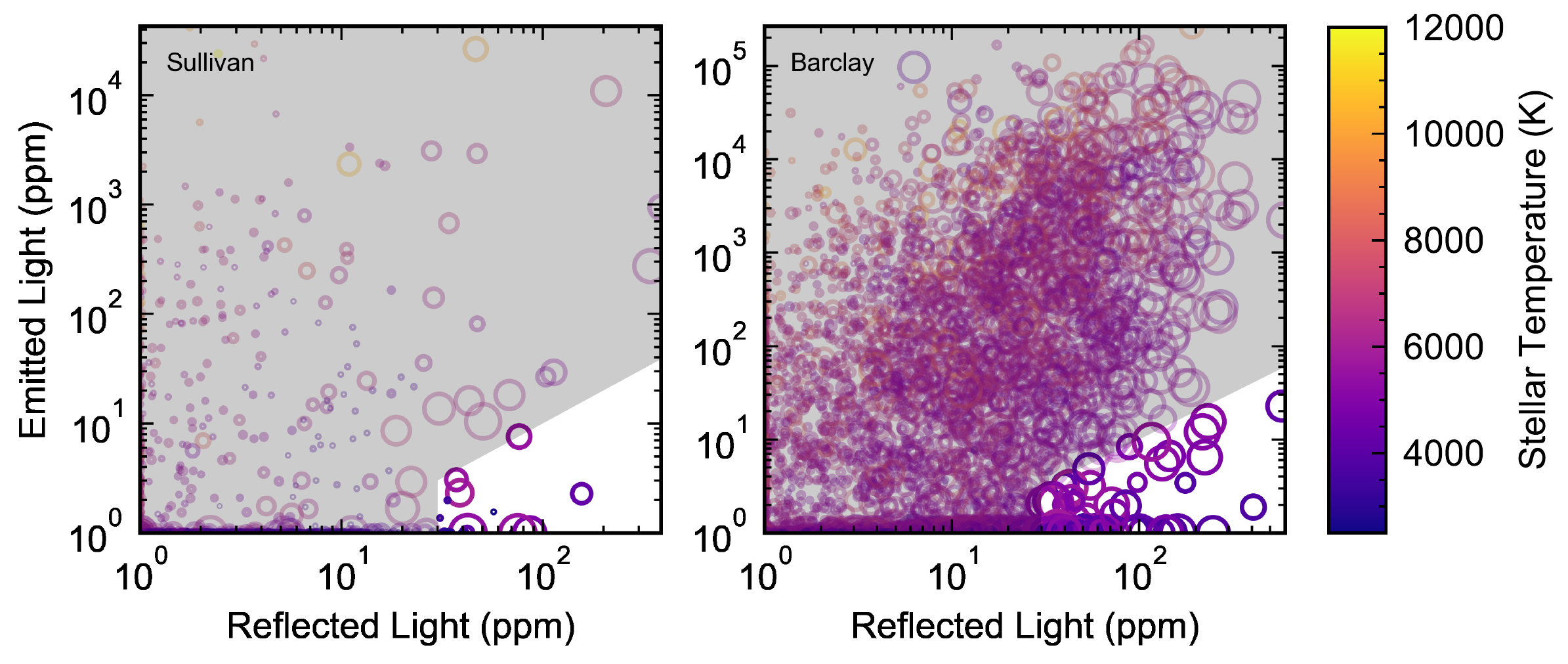}
    \caption{The reflected light vs emitted light of the \citet{Sullivan2015} predicted planets (left) and the \citet{Barclay2018} predicted planets (right). The color indicates the effective temperature of the host star and the symbol size scales with the radius of the planet. The selected sample occupies the unshaded region, which denotes the region matching our criteria of a large enough reflected light signal for detectability with little to no thermal contamination in the \tess{} bandpass. \label{fig:refvem}}
\end{figure*}

We first estimate the emitted light of the predicted planets by assuming both the planet and the star are blackbodies ($B_{\lambda}(T_{\rm eq}$ and $B_{\lambda}(T_{\rm eff}$)and then observe the system using the \tess{} bandpass,
\begin{equation}
    F_{\rm E} = \frac{\int B_{\lambda}(T_{\rm eq}) T_\lambda{\rm d}\lambda}{\int T_\lambda {\rm d} \lambda},
\end{equation}
where $T_{\lambda}$ is the transmission curve of the bandpass in question, before determining their ratio $F_{\rm E}/F_*$. While \added{neither} the planet \replaced{and}{nor the} star are \deleted{not} blackbodies, this is a useful diagnostic for sample selection and we return to these assumptions in \autoref{sec:results}. The computed reflected and emitted light of the planets is shown in \autoref{fig:refvem}. The sample of available planets is greatly reduced by this requirement due to the \tess{} bandpass being far redder than the bandpasses of previous space-based exoplanet surveys.

The aforementioned selection criteria reduce the sample to 81 planets from the \citet[][Arxiv version 1]{Barclay2018} sample and 21 from the \citet{Sullivan2015} sample. The planets are generally around cool stars ($<$5500~K) with short orbital periods ($<$9~days) and all are cooler than 1200~K. The redder bandpass of \tess{} causes this resulting sample to be smaller than any similar sample selection from a survey with a bluer bandpass because of the increased risk of thermal contamination. With \tess{} the selected sample is only 1.05\% of \citet{Sullivan2015} and 1.78\% of \citet{Barclay2018}. Using the \kepler{} bandpass these increase to 1.36\% and 3.07\% respectively and with a box filter from 0.3--0.5$\mu$m these increase to 1.56\% and 7.75\%.

However, there are a number of planets in the datasets which are actually phantom inflated planets \citep[see][\added{for a deeper explanation of the issue}]{Mayorga2018RNAAS, Barclay2018}, assumed to have radii reaching almost 2 $R_J$ with equilibrium temperatures below 1200~K. Previous work has shown that planets larger than 1.25~$R_J$ are essentially prohibited at temperatures below $\sim$1000~K unless they are very young \citep{Thorngren2016, Thorngren2018}. We apply a conservative inflation limit cut-off based on \citet{Thorngren2018} to eliminate inflated planets in the sample that were larger than 1.2~$R_J$+$\Delta R_J$, where $\Delta R_J$ is the additional radius added from inflation as function of temperature as calculated therein. This leaves 20 planets from the \citet{Barclay2018} sample and 17 from the \citet{Sullivan2015} sample.

\subsection{The Representative Sample: The Nine}
\begin{figure*}
    \plotone{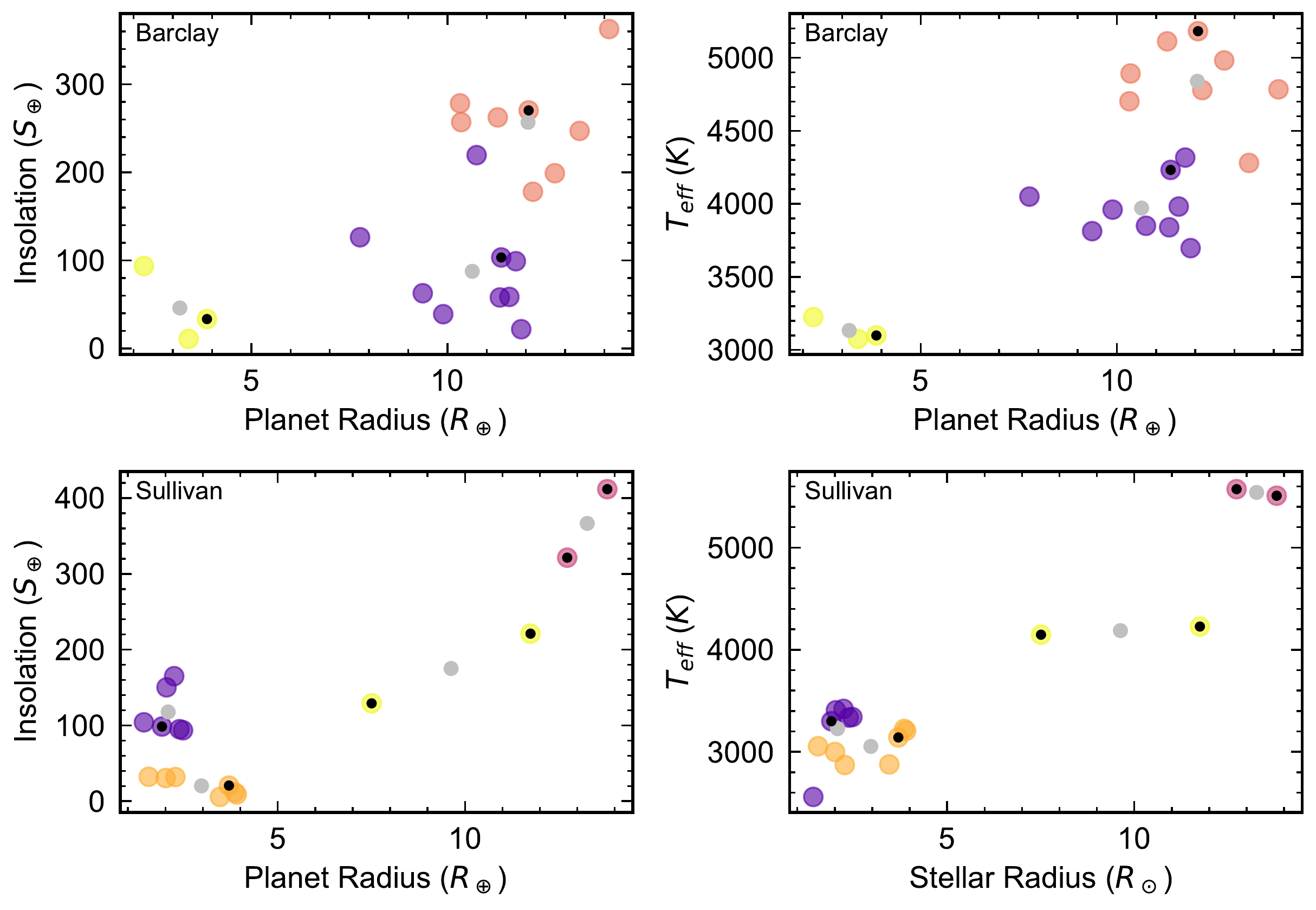}
    \caption{The results of the k-means clustering algorithm with three clusters for the \citet{Barclay2018} sample (top) and four clusters for the \citet{Sullivan2015} sample (bottom). The clusters are color coded by cluster membership and the cluster centers are shown in silver. In black are the positions of the nine representatives we chose for modeling.\explain{This figure has also been updated. We identified after submission that the y label on the Barclay Teff vs Radius plot incorrectly read P (days)}\label{fig:kmeans}}
\end{figure*}

To generate an artificial representative sample of planets that may be found by \tess{}, we down-selected to a more manageable number of planets through the use of a k-means clustering algorithm \citep{sklearn}. We clustered the set of planets predicted by both works according to the parameters of planetary radius, stellar effective temperature, and insolation. From the distortion and silhouette analyses we concluded that the \citet{Barclay2018} sample was best represented with three clusters and the \citet{Sullivan2015} was best represented with four. The results of the clustering algorithm are shown in \autoref{fig:kmeans}. 

Instead of recomputing all other planet properties for the computed cluster center, we determine the closest member to be the representative. However, for the \citet{Sullivan2015} sample, the giant planet population is small and the clusters have only two members each. Therefore, we chose all four giant planets as representatives for further modeling. Thus, six planets were chosen from that sample leading to a total of nine planets to model hereafter referred to as The Nine.

For each of The Nine, we go on to compute the planets' mass and gravity, and the stars' luminosities and gravity. To generate planet masses, we use the relation given by \citet{Weiss2013} derived from a sample of 138 planets whose masses, radii, and orbital semi-major axes were measured with errors given and stellar temperatures and radii were measured with errors given. For planets with $M_P < 150~M_\oplus$,
\begin{equation}
\frac{R_{\rm p}}{R_\oplus} = 1.78\left(\frac{M_{\rm p}}{M_\oplus}\right)^{0.53}\left(\frac{F}{\rm ergs\,s^{-1}\,cm^{-2}}\right)^{-0.03},
\end{equation}
\added{
and for planets $M_P > 150~M_\oplus$,
\begin{equation}
\frac{R_{\rm p}}{R_\oplus} = 2.45\left(\frac{M_{\rm p}}{M_\oplus}\right)^{-0.039}\left(\frac{F}{\rm ergs\,s^{-1}\,cm^{-2}}\right)^{0.094},
\end{equation}
}
where we assume that the incident flux F = $8.6\times10^8$ ergs s$^{-1}$ cm$^{-2}$ for all planets \citep[the median flux from][]{Weiss2013} \added{to be agnostic about the inflated or non-inflated status of the planets in the samples}. 

\begin{figure*}
    \plotone{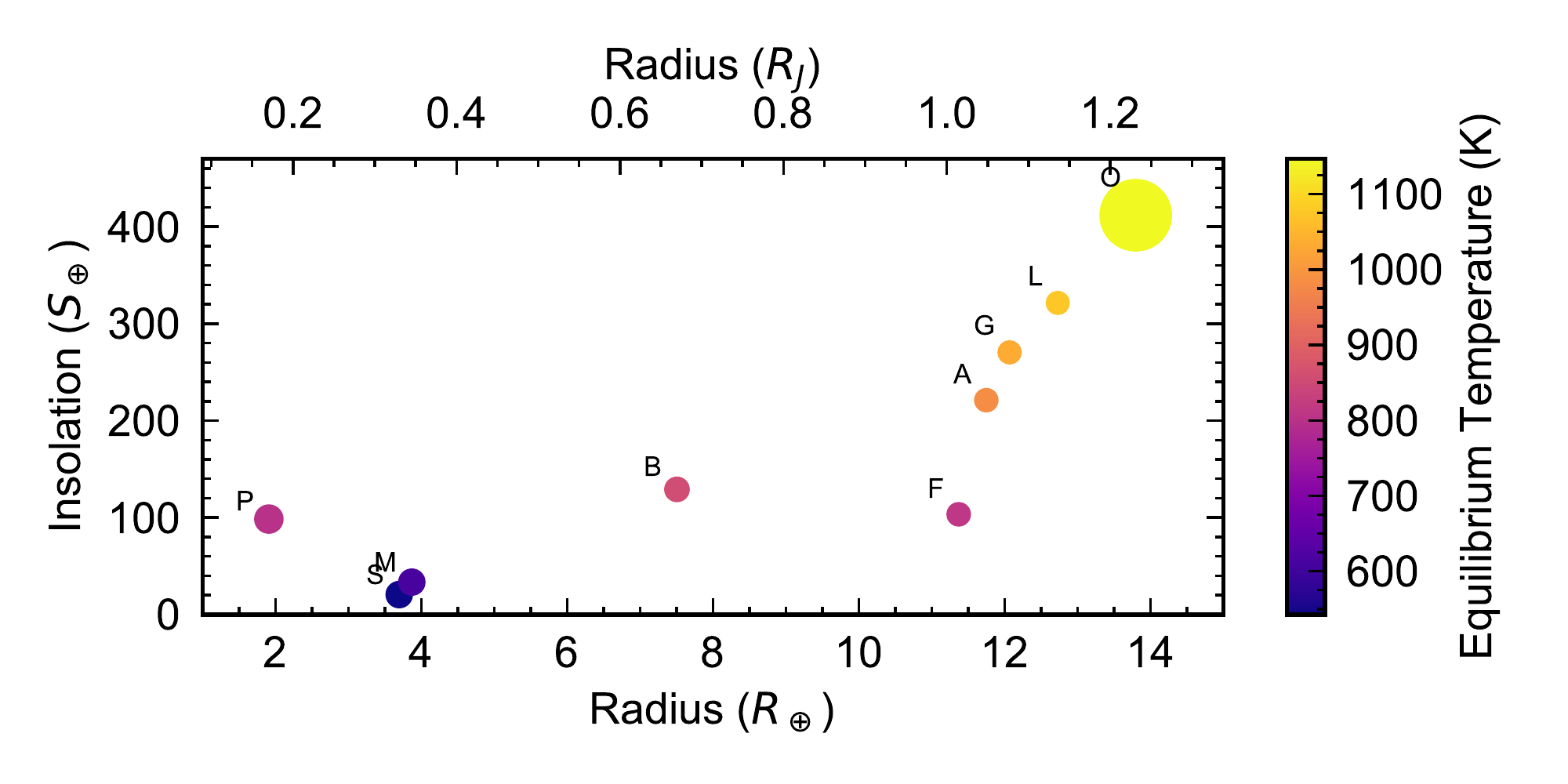}
    \caption{The Nine labeled with their associated letter. The symbol size is proportional to the planet's mass and the color with the computed equilibrium temperature.\label{fig:reps}}
\end{figure*}

The Nine are members of a variety of planet classes. The properties of the representative sample are shown in \autoref{tbl:reps} and in \autoref{fig:reps}. Planet P is the smallest planet and is the closest to its host star and has the shortest period, S and M are similar in radius and mass to Neptune and are the coolest of the Nine. These three planets have the coolest host stars. Planet B is larger than Neptune. The remaining planets track a series of warmer and warmer Jupiter-class planets and are around the hottest host stars. Planet F is the longest period planet at just under 4 days. Planet A is the closest to its host star of the Jupiter-class planets. Planet L has the lowest gravity and is furthest from its host star. Planet O has the highest gravity and is the hottest.

\begin{deluxetable*}{LRRRRRRRRR}
\tablecaption{Table of properties for the representative planets and their stellar hosts.\label{tbl:reps}}
\tablehead{\colhead{Name} & \colhead{$P$} & \colhead{$a$} & \colhead{$R_{\rm P}$} & \colhead{$M_{\rm P}$} & \colhead{$T_{\rm eq}$} & \colhead{$S$} & \colhead{g} & \colhead{\hspace{2em}$T_{\rm eff}$} & \colhead{$\log g$}\\ \colhead{} & \colhead{days} & \colhead{AU} & \colhead{$R_\oplus$} & \colhead{$M_\oplus$} & \colhead{$K$} & \colhead{$S_\oplus$} & \colhead{m\,s$^{-2}$} & \colhead{\hspace{1.5em}$K$} & \colhead{log(cm\,s$^{-2}$)}}
\startdata
P & 0.515 & 0.008 & 1.9 & 3.7 & 802 & 98 & 9.835 & 3300 & 5.017 \\
S & 2.027 & 0.015 & 3.7 & 12.7 & 542 & 20 & 9.123 & 3141 & 4.957 \\
M & 0.920 & 0.011 & 3.9 & 13.9 & 612 & 33 & 9.076 & 3099 & 5.052 \\
B & 2.070 & 0.027 & 7.5 & 48.4 & 858 & 129 & 8.419 & 4147 & 4.627 \\
F & 3.948 & 0.046 & 11.4 & 106.0 & 811 & 103 & 8.033 & 4232 & 4.482 \\
A & 1.488 & 0.022 & 11.8 & 112.8 & 981 & 221 & 8.003 & 4227 & 4.621 \\
G & 3.328 & 0.043 & 12.1 & 118.6 & 1032 & 270 & 7.979 & 5181 & 4.527 \\
L & 3.913 & 0.049 & 12.7 & 131.2 & 1077 & 321 & 7.931 & 5572 & 4.464 \\
O & 2.738 & 0.037 & 13.8 & 499.6 & 1146 & 411 & 25.705 & 5508 & 4.543 \\
\enddata
\end{deluxetable*}

\section{Models}
\label{sec:models}
For each planet, we generate \replaced{1D}{one dimensional (1D) atmospheric} structure models \added{appropriate for hydrogen-helium dominated atmospheres} and then compute their albedo spectra. \added{Of course since these are models of hypothetical planets there is substantial uncertainty, which we capture by exploring a  range of model parameters.} The 1D structures are computed using the irradiated giant planet atmospheres code of \citet{Marley1999b} which is based on \citet{McKay1989} \citep[see also][]{Marley2015}. The albedo spectra are computed based on \citet{Marley1999} as modified by \citet{Cahoy2010} and Batalha et al. (2019) to handle arbitrary phase observations. Raman scattering was updated in Batalha et al. (2019) to include Raman ghost features \citep{Oklopcic2016}, but we utilize the original \citet{Pollack1986} methodology because it retains the overall dampening of reflectivity toward the blue (important for photometric observations).

\begin{figure}
    \plotone{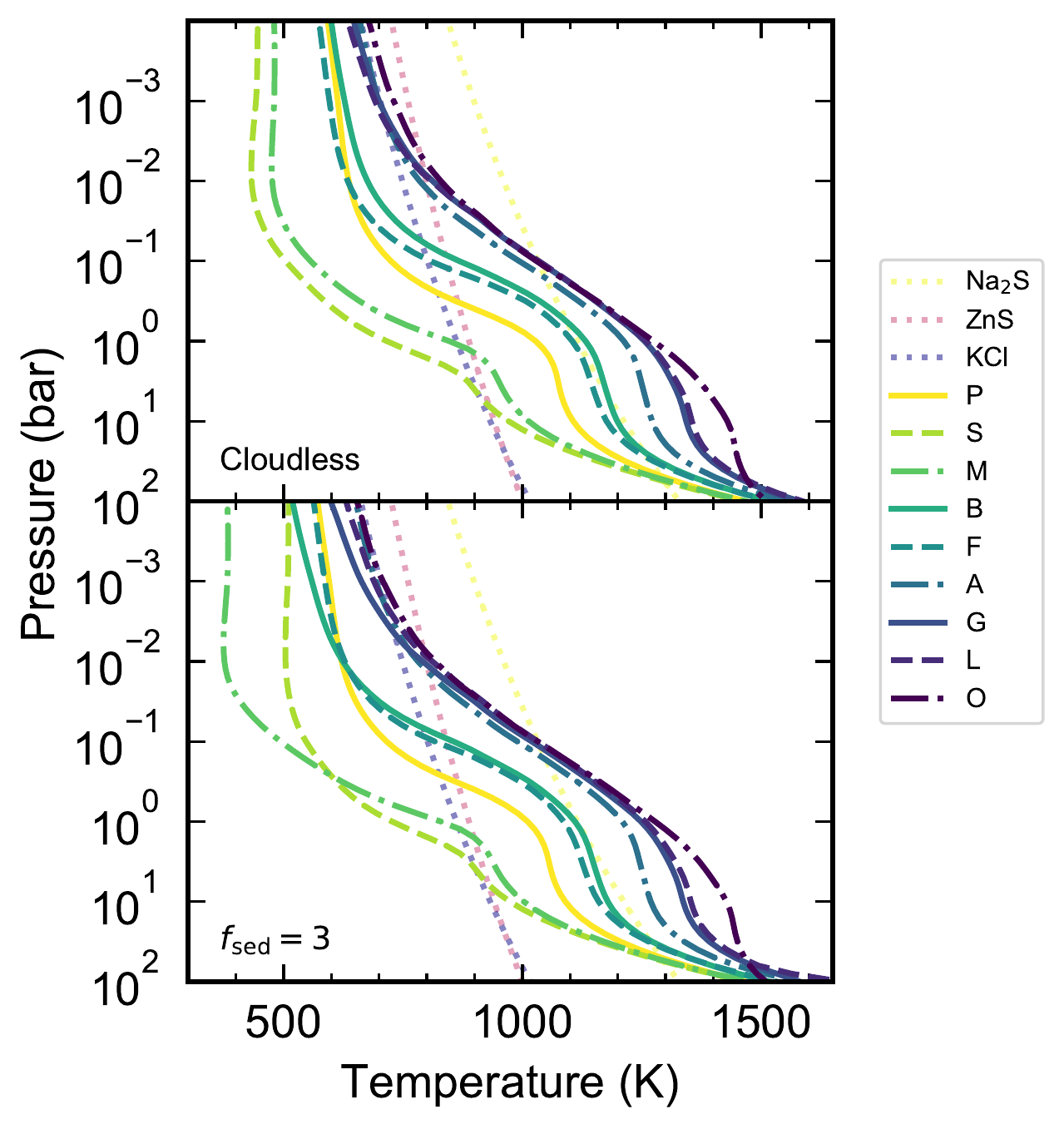}
    \caption{The temperature pressure profiles for all nine planets for the cloudless (top) and cloudy $f_{\rm sed}$=3 (bottom) cases. For demonstration, we show the condensation curves of the modeled cloud species in dashed lines \citep{Morley2012}. \label{fig:cloudcomp}}
\end{figure}

We first compute cloud free models by generating an atmospheric radiative-convective equilibrium thermal profile for each planet assuming an internal heat flux, $\sigma T_{\rm int}^4$, with $T_{\rm int} = 150\,\rm K$ and assuming solar abundances for both the planets and their stars. We also computed self-consistent cloudy profiles for each planet while setting the cloud parameter to $f_{\rm sed}$=3 \citep{Ackerman2001}. The $f_{\rm sed}$ parameter controls the particle size and vertical extent of the cloud layer. A small $f_{\rm sed}$ yields tall lofted clouds of small particles while a large $f_{\rm sed}$ produces a vertically thin cloud with generally larger particles. The following gas species were allowed to condense: Al$_2$O$_3$, Fe, MgSiO$_3$, Cr, MnS, Na$_2$S, ZnS, KCl, H$_2$O, and NH$_3$ \citep{Morley2012, Marley2013}.

The temperature pressure profiles for the cloudless and $f_{\rm sed}$=3 cases are shown in \autoref{fig:cloudcomp}. Typically reflected light probes depths down to 1~bar, but light may penetrate as deep as 10~bar depending on where the atmosphere becomes opaque. \replaced{The models}{Model atmospheric temperatures} range from several hundred Kelvin at pressures of a millibar to as hot as $\sim$1400~K at 10 bars. In the region from 10$^{-4}$~bar to 10~bars, the condensing species are Na$_2$S, ZnS, KCl. The others will either not condense because the atmosphere is too hot, such as H$_2$O, and NH$_3$, or because they have already condensed deeper in the atmosphere. In \autoref{fig:cloudcomp}, we show the condensation curves for the potential gas species for the pressure range in question. 

The cloud and thermal structure computed for the $f_{\rm sed}$=3 case is given to the albedo code to compute the reflected light spectrum. To facilitate comparison between different $f_{\rm sed}$ cases, we use the same thermal profile for the other $f_{\rm sed}$ cases, with only the cloud opacity varied for each case similar to the methodology of \citet{Batalha2018Color, Morley2015}. We generate albedo spectra for cloud free atmospheres and atmospheres with $f_{\rm sed}$=3, 1, 0.3, and 0.1. These span a range from very extensive clouds to thin cloud decks. Most brown dwarfs have been fit with clouds in a range of $f_{\rm sed}$=1 to $f_{\rm sed}$=3 \citep{Stephens2009} and \cite{Ackerman2001} found $f_{\rm sed}$=3 reproduces Jupiter's spectrum. Smaller $f_{\rm sed}$ were chosen to account for extremely extended clouds such as have been proposed for hot Jupiters \citep{Demory2013, Webber2015} and super-Earths \citep{Morley2013, Morley2015}.

\added{The range of models resulting from this procedure is meant to span a range of plausible cases, not encompass every possible extrasolar planet with the selected mass. In particular at lower masses the mass-metallicity trend (e.g., \citealp{Wakeford2018}) tentatively indicates that the atmospheres of lower mass planets are generally more enriched in heavy elements than higher mass planets. In reflected light the effect of metallicity variations is not always intuitive. In a study of the behavior of water opacity in reflected light giant planets, \citet{Macdonald2018} found that as metallicity increased cloud height also increased, resulting in weaker $\rm H_2O$ absorption bands and brighter planets. We used the grid of reflected light planets from \citet{Batalha2018Color} to do a metallicity sensitivity analysis on reflectivity. In general, metallicity worked to increase the total opacity of the atmospheres and ultimately decreased the reflectivity of the planets. Limiting ourselves to solar metallicity cases presents an optimistic view of how bright the planets could be.}

\added{Another important uncertainty is the neglect of UV absorbers. In the Solar System many planets, from Venus, to Neptune, exhibit reduced reflectivity at UV wavelengths \citep[e.g.,][]{Pollack1979,Savage1980}, often arising from the presence of disequilibrium gas and haze species resulting from photochemistry. Indeed S-bearing hazes expected in cool hydrogen dominated atmospheres such as those considered here can dramatically lower the UV albedo while brightening the planet at redder wavelengths \citep{Gao2017}. Since the present work is primarily concerned with understanding relative reflectivity as a function of wavelength and the degree to which thermal emission contaminates optical bandpasses over the selected portion of phase space, we neglect such possibilities. Any interpretation of observed planets would of course need to consider such mechanisms as these.}

\added{The modeling suite used in this study has been applied to numerous studies of atmospheric structure and spectroscopic signatures in the solar system (such as \citealp{McKay1989, Marley1999b, Cahoy2010}; etc), it has been applied to brown dwarfs (such as \citealp{Marley1996, Burrows1997, Robinson2014}; etc.), super-Earths (see \citealp{Morley2013, Morley2015}; etc), and giant exoplanets (such as \citealp{Fortney2005, Fortney2008}; etc) including the interpretation of the optical phase variations of  Kepler-7b \citep{Demory2013, Webber2015}.}

\section{Results/Discussion}
\label{sec:results}
\subsection{Albedo}
\begin{figure*}
    \centering
    \includegraphics[height=0.8\textheight]{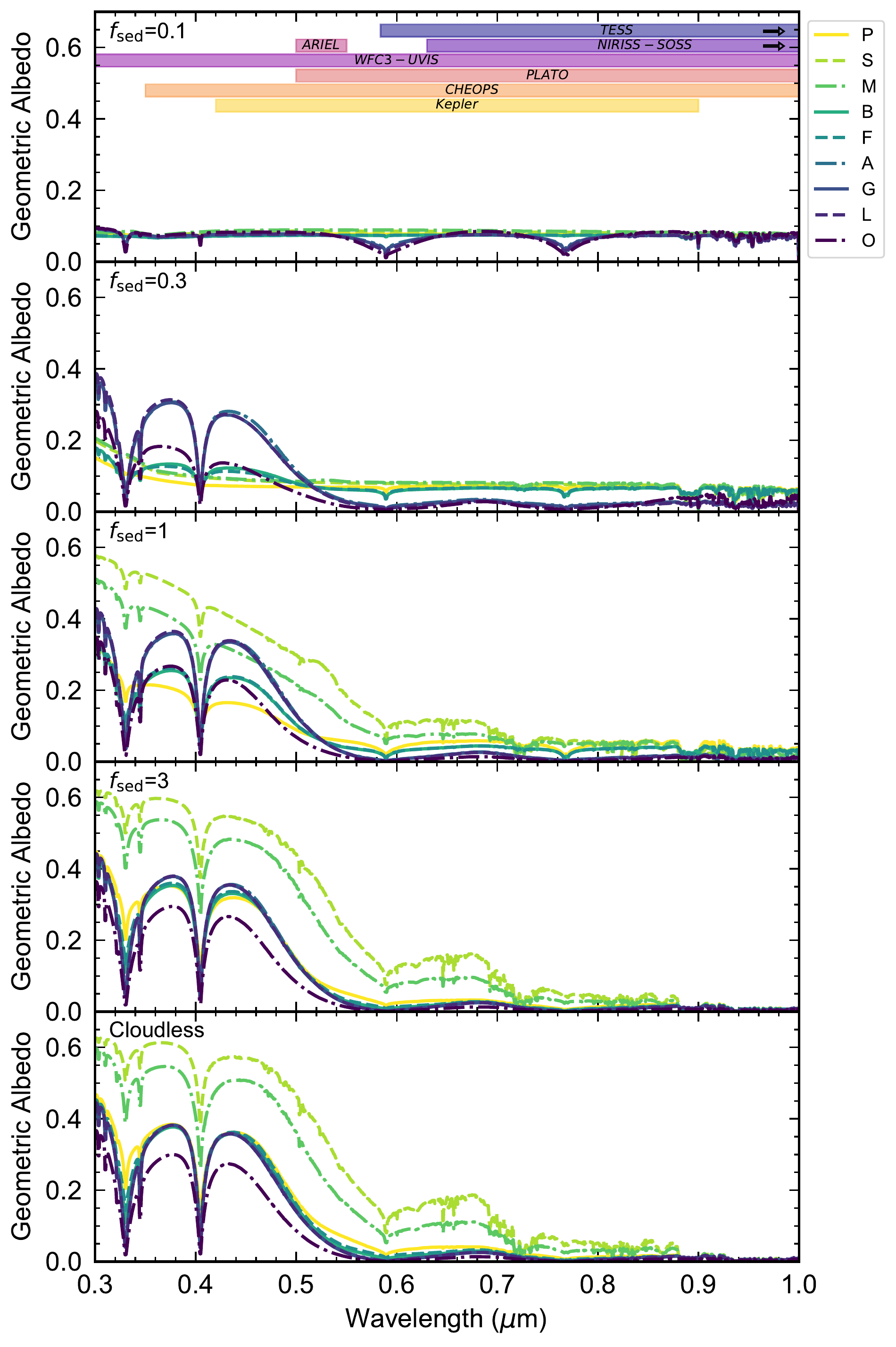}
    \caption{Model geometric albedo spectra for The Nine with varying cloudiness and $f_{\rm sed}$. The colored bars indicate the bandpasses or wavelength coverage of present and future space-based missions for exoplanet discovery and characterization \added{\citep{Ricker2014,ARIEL,SOSS,PLATO2,Broeg2013,Borucki2010}}. Planets like these will have very little reflected light signal in the region predominantly shared by future missions. \label{fig:albedo}}
\end{figure*}

\added{With the assumptions we have made above we find that,} in general, The Nine have a small reflected light signal in the \tess{} bandpass regardless of assumed $f_{\rm sed}$\replaced{ and}{Additionally, } this will likely also be the case for \added{observations made of the Nine and the planets they embody with }\plato{} \added{\citep{PLATO2}}, \jwst{}, and \added{in} all but perhaps the shortest channel of \ariel{} \added{\citep{ARIEL}}. 
This is consistent with preliminary \tess{} results such as \citet{Shporer2019}. The predicted geometric albedo spectra for The Nine are shown in \autoref{fig:albedo}. The tabulated geometric albedos in the \tess{} bandpass are shown in \autoref{tbl:geomalb} and we also considered a strawwoman blue filter, a box filter from 0.3--0.5~$\mu$m, for comparison.

\begin{deluxetable*}{lRRRRRRRRRR}
\tabletypesize{\small}
\tablecaption{Geometric albedo of the planets in different bandpasses\tablenotemark{a}.\label{tbl:geomalb}}
\tablehead{\multirow{2}{*}{Planet} & \multicolumn{2}{c}{$f_{\rm sed}$=0.1} & \multicolumn{2}{c}{\hspace{1.5em}$f_{\rm sed}$=0.3} & \multicolumn{2}{c}{\hspace{1.5em}$f_{\rm sed}$=1} & \multicolumn{2}{c}{\hspace{1.5em}$f_{\rm sed}$=3} & \multicolumn{2}{c}{\hspace{1.5em}Cloudless}\\ & \colhead{$A_{\rm blue}$} & \colhead{$A_\tess{}$} & \colhead{\hspace{1.5em}$A_{\rm blue}$} & \colhead{$A_\tess{}$} & \colhead{\hspace{1.5em}$A_{\rm blue}$} & \colhead{$A_\tess{}$} & \colhead{\hspace{1.5em}$A_{\rm blue}$} & \colhead{$A_\tess{}$} & \colhead{\hspace{1.5em}$A_{\rm blue}$} & \colhead{$A_\tess{}$}}
\startdata
P & 0.07 & 0.08 & 0.09 & 0.07 & 0.18 & 0.04 & 0.30 & 0.02 & 0.33 & 0.02 \\
S & 0.08 & 0.08 & 0.11 & 0.07 & 0.45 & 0.06 & 0.55 & 0.06 & 0.57 & 0.07 \\
M & 0.08 & 0.08 & 0.12 & 0.07 & 0.36 & 0.05 & 0.48 & 0.04 & 0.49 & 0.04 \\
B & 0.07 & 0.07 & 0.12 & 0.06 & 0.21 & 0.03 & 0.28 & 0.01 & 0.30 & 0.01 \\
F & 0.07 & 0.07 & 0.11 & 0.06 & 0.21 & 0.03 & 0.29 & 0.01 & 0.31 & 0.01 \\
A & 0.08 & 0.07 & 0.24 & 0.02 & 0.28 & 0.01 & 0.29 & 0.01 & 0.30 & 0.01 \\
G & 0.08 & 0.07 & 0.23 & 0.02 & 0.28 & 0.01 & 0.29 & 0.01 & 0.29 & 0.01 \\
L & 0.08 & 0.07 & 0.24 & 0.02 & 0.28 & 0.01 & 0.29 & 0.01 & 0.29 & 0.01 \\
O & 0.08 & 0.07 & 0.13 & 0.02 & 0.19 & 0.01 & 0.21 & 0.01 & 0.22 & 0.01
\enddata
\tablenotetext{a}{The blue bandpass is a simple top-hat filter from 0.3--0.5$\mu$m.}
\end{deluxetable*}

The largest geometric albedos are of order 0.1 in the \tess{} bandpass for the lowest sedimentation efficiency cases i.e. $f_{\rm sed}$=0.1 and some planets with $f_{\rm sed}$=0.3. The cloudless and higher sedimentation efficiency cases, universally, have geometric albedos of order 0.05 or less.  Planets S and M boast the highest albedos in the \tess{} bandpass followed by P, B, and F. These are the smaller planets in the Nine. The larger planets are all consistently dark.

\begin{figure*}
    \plotone{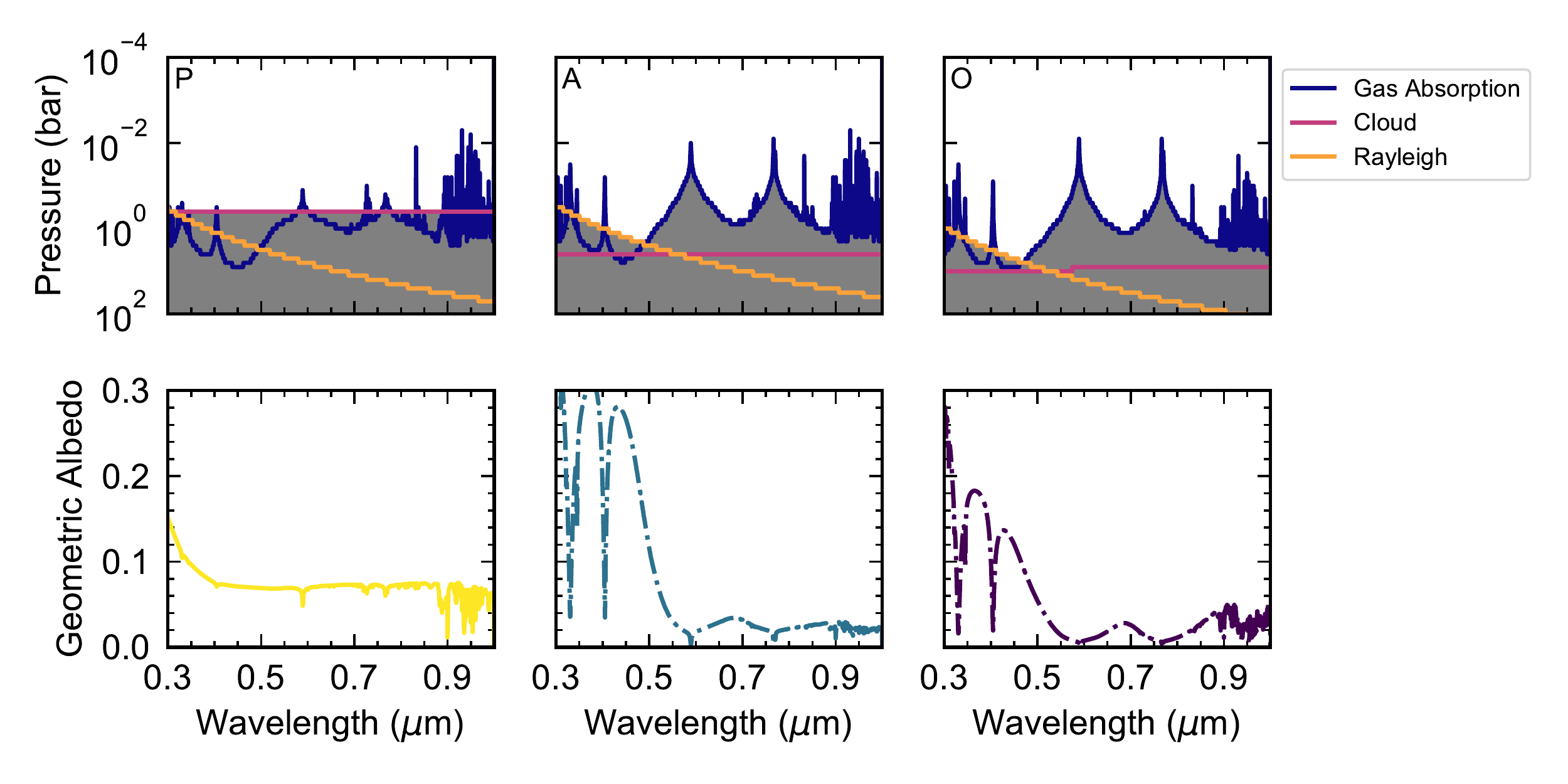}
    \caption{\replaced{The p}{P}hoton attenuation plot\added{s} for \added{planets P, A, and O} \replaced{the}{with} $f_{\rm sed}$=0.3\deleted{case of planets P, A, and O}. Top: the pressure at which the gas opacity reaches $\tau$=1 (blue), the cloud opacity reaches $\tau$=1 (pink), and the Rayleigh scattering reaches $\tau$=1 (orange). The incident light is essentially probing the atmosphere above the pressure level at which the atmosphere has become \replaced{opaque}{optically thick} (a combination of all three opacity sources)\explain{The figure has also been updated}. We have shaded the deeper and unobservable pressures in gray. Bottom: the geometric albedo spectrum for the same three planets. Note how the shapes are reflective of the gas opacity at long wavelengths and the Rayleigh scattering slope at short wavelengths when clouds do not obstruct the view of deeper pressures. \label{fig:atmotrans}}
\end{figure*}

Because the clouds are mostly gray, the pressure level at which they become opaque is mostly independent of wavelength. In the $f_{\rm sed}$=0.1 case where the clouds dominate the opacity the albedos of the planets are similar in the blue and the albedo spectrum is nearly flat in \autoref{fig:albedo}. With increasing sedimentation efficiency, the albedos in the blue actually increase because, instead of being dominated by gas absorption, Rayleigh scattering is keeping the planets bright. The transition at $f_{\rm sed}$=0.3 causes the larger, low gravity planets (A, G, and L) to be significantly brighter than the rest in this regime. At higher sedimentation efficiencies, planets S and M are brighter, likely because these are the coolest planets in the Nine. The lowest albedo is typically planet O, the largest, warmest, and highest gravity planet of the Nine, except in the transition regime at $f_{\rm sed}$=0.3 when planet P, the smallest, is the darkest.

These results can be understood as a competition between gas and cloud scattering and gaseous absorption. In cloudless cases incoming photons are either Rayleigh scattered or absorbed. Absorption dominates over Rayleigh scattering in the red, leading to low reflectivity while in the blue scattering is more important and the planets have higher albedos. The addition of clouds provides a new source of scattering opacity, brightening the albedos in the red, for example raising $A_{TESS}$ from 0.02 to 0.08 for Planet P at $f_{\rm sed}=0.1$. However the main source of cloud opacity at these wavelengths is $\rm Na_2S$, a condensate which can be dark at optical wavelengths \citep[see][]{Morley2015} with single scattering albedos near 0.6, depending on cloud particle size. This has the effect of lowering the geometric albedo in the blue in cases where the cloud opacity dominates Rayleigh scattering, lowering Planet P's albedo in the blue from 0.33 to 0.07 in this same case.

Indeed the albedo spectra in the low sedimentation efficiency regime are nearly featureless in \autoref{fig:albedo} because the high clouds obscure almost all of the gas below them, absorbing and scattering the light before any gas absorption can occur. This is true for all but the strongest absorption features, such as the Na and K lines, where the absorption of that particular wavelength occurs much higher in the atmosphere than where the clouds would make the atmosphere opaque. At higher sedimentation efficiencies the cloud decks are lower and less vertically extensive so they do not contribute to the sources of opacity in the upper atmosphere. Instead, the gas causes the atmosphere to become opaque above the cloud layer and the albedo spectra have more absorption line features, dominated by Na and K in the blue with contribution from CO$_2$ and H$_2$O at redder wavelengths.

The interplay between the gas opacity via absorption, the cloud opacity, and Rayleigh scattering is presented in \autoref{fig:atmotrans} where we show three photon attenuation plots for three planets with $f_{\rm sed}$=0.3 atmospheres. A photon attenuation plot shows the pressure level at which the two-way optical depth in the atmosphere reaches $\tau$=1. The planets get darker with increasing sedimentation efficiency as the gas opacity becomes the dominant source of opacity rather than the cloud opacity, becoming equivalent to the cloudless case. This transition from cloud dominated opacity to gas dominated opacity, from Rayleigh scattering and absorption, begins to occur in some of the planets at $f_{\rm sed}$=0.3 (see \autoref{fig:albedo}).

\subsection{Contrast}

\begin{figure*}
\centering
    \includegraphics[height=0.8\textheight]{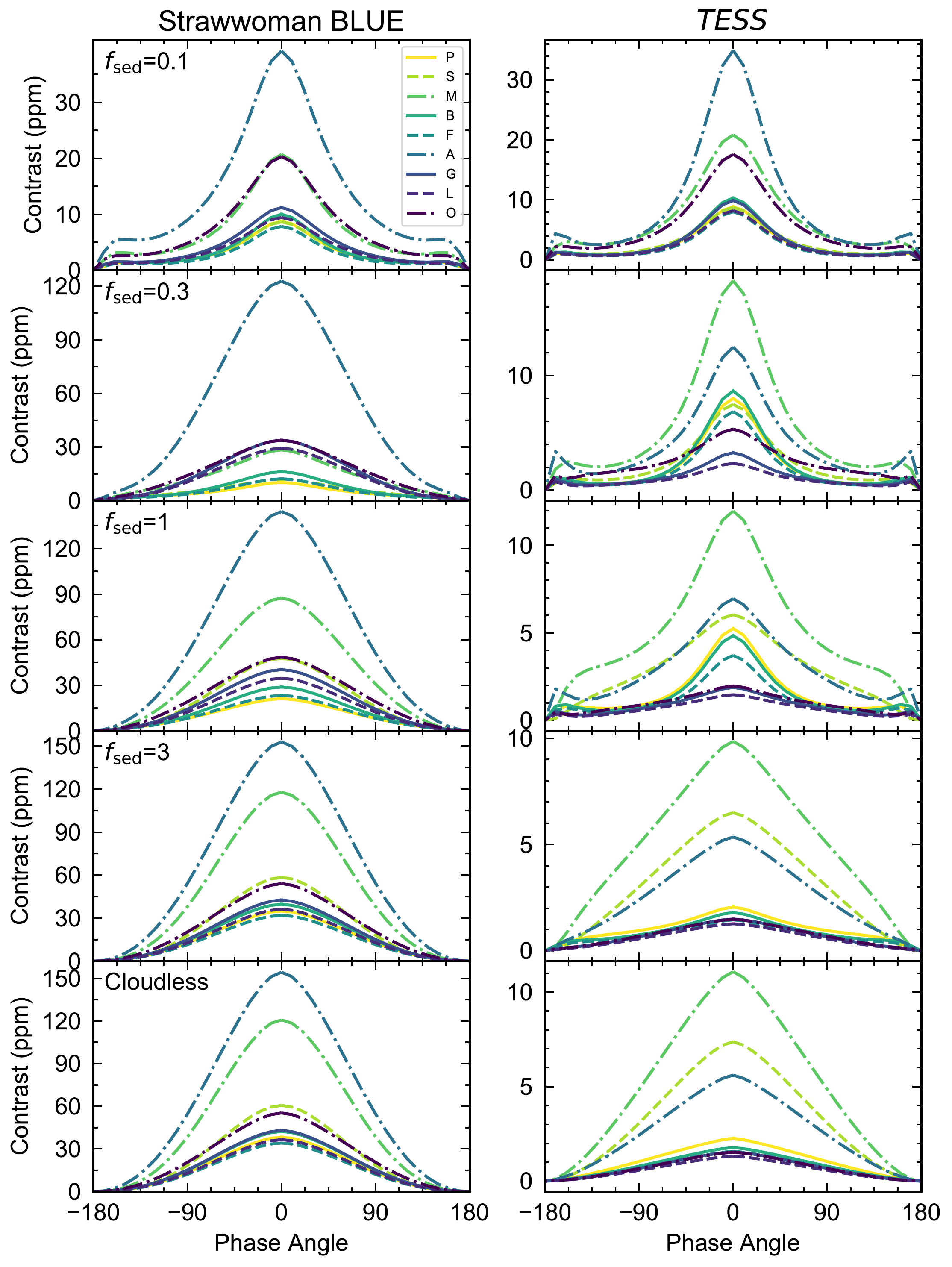}
    \caption{The contrast phase curves for The Nine as a function of cloudiness and $f_{\rm sed}$ in a strawwoman blue filter (left) and the \tess{} bandpass (right) assuming uniform cloud coverage. \label{fig:contrast}}
\end{figure*}

We compute the contrast ratio, $C$, for the representative sample using the formalism given in \citet{Cahoy2010},
\begin{equation}
C=A_{\rm g}\left(\frac{R_{\rm p}}{a}\right)^2\Phi(\alpha),
\end{equation}
where $\Phi(\alpha)$ is the phase curve (the reflectivity of the planet as a function of phase angle, $\alpha$, as computed by the albedo code) for the cloudless case, all cloudy cases, and in both the \tess{} bandpass and the strawwoman blue filter. The contrast curves are shown in \autoref{fig:contrast} in parts-per-million (ppm) and \added{the maximum contrasts are} tabulated in \autoref{tbl:contrast}. For simplicity, we assume a uniform cloud distribution \added{across the planet}. Deviations and patchy cloud coverage would cause additional features in the phase curve.

\begin{deluxetable*}{lRRRRRRRRRR}
\tabletypesize{\small}
\tablecaption{Planet contrast in ppm for different bandpasses at full phase.\label{tbl:contrast}}
\tablehead{ \multirow{2}{*}{Planet} & \multicolumn{2}{c}{$f_{\rm sed}$=0.1} & \multicolumn{2}{c}{\hspace{1.5em}$f_{\rm sed}$=0.3} & \multicolumn{2}{c}{\hspace{1.5em}$f_{\rm sed}$=1} & \multicolumn{2}{c}{\hspace{1.5em}$f_{\rm sed}$=3} & \multicolumn{2}{c}{\hspace{1.5em}Cloudless} \\  & \colhead{$C_{\rm blue}$} & \colhead{$C_\tess{}$} & \colhead{\hspace{1.5em}$C_{\rm blue}$} & \colhead{$C_\tess{}$} & \colhead{\hspace{1.5em}$C_{\rm blue}$} & \colhead{$C_\tess{}$} & \colhead{\hspace{1.5em}$C_{\rm blue}$} & \colhead{$C_\tess{}$} & \colhead{\hspace{1.5em}$C_{\rm blue}$} & \colhead{$C_\tess{}$}}
\startdata
P & 8.8 & 8.8 & 10.2 & 8.0 & 21.2 & 5.2 & 35.0 & 2.1 & 38.3 & 2.3 \\
S & 8.7 & 8.7 & 12.0 & 7.5 & 47.8 & 6.0 & 58.5 & 6.5 & 60.6 & 7.4 \\
M & 20.7 & 20.8 & 28.4 & 18.3 & 87.5 & 12.0 & 117.8 & 9.8 & 120.7 & 11.1 \\
B & 10.0 & 10.3 & 16.2 & 8.7 & 28.9 & 4.8 & 39.7 & 1.8 & 42.3 & 1.8 \\
F & 7.8 & 8.0 & 12.2 & 6.9 & 23.3 & 3.7 & 32.0 & 1.5 & 34.2 & 1.6 \\
A & 39.2 & 34.9 & 122.8 & 12.5 & 144.3 & 7.0 & 152.8 & 5.3 & 154.3 & 5.6 \\
G & 11.2 & 9.9 & 33.8 & 3.3 & 40.3 & 1.9 & 42.7 & 1.5 & 43.1 & 1.5 \\
L & 9.4 & 8.2 & 29.3 & 2.3 & 34.7 & 1.5 & 36.2 & 1.3 & 36.5 & 1.3 \\
O & 20.3 & 17.6 & 33.8 & 5.3 & 48.5 & 2.0 & 54.1 & 1.5 & 55.3 & 1.5
\enddata
\end{deluxetable*}

The albedos in the \tess{} bandpass are so small that they have little correlation with the predicted contrasts. The contrasts are instead controlled by planet size and orbital semi-major axis, the strongest dependencies in the contrast equation. Since the albedos decrease with increasing $f_{\rm sed}$ the contrasts do the same. The computed phase curve amplitudes in \tess{} observations range between 1~ppm in the cloudless case and increase with decreasing sedimentation efficiency up to as much as $\sim$35~ppm. Since the geometric albedos in the $f_{\rm sed}$=0.1 case are roughly the same for each planet, planet A is the brightest due to it being the largest planet that is closest to its host star of the Nine. Otherwise, in other sedimentation efficiency cases, planet M is the brightest because it is slightly larger than planet S and an almost comparable high albedo in the \tess{} bandpass.

In general, the \replaced{planets}{Nine} are much brighter in the blue than in the \tess{} bandpass ranging from 8~ppm in the $f_{\rm sed}$=0.1 case and increasing with $f_{\rm sed}$ to tens of ppm and for planet A to over 150~ppm. At higher sedimentation efficiencies planets S and M had the highest albedos in the blue, planet A's proximity to its host star continues to be the dominating factor in making the planet have such a large contrast ratio. Planet M is a close second at all sedimentation efficiencies.

Additionally, the shape of the phase curve is different between the two bandpasses. The blue bandpass shows a \deleted{sinusoidal} phase curve\added{, reminiscent of the classic Lambertian or sinusoidal phase curve assumption,} dominated by Rayleigh scattering\replaced{, with t}{. T}he exception \replaced{of}{is the} $f_{\rm sed}$=0.1 \added{scenario in the blue bandpass} which \replaced{tracks more closely with the shapes of }{is similar to} the particle-dominated phase curves \added{seen} in the \tess{} bandpass\replaced{. T}{; t}hese phase curves show much structure, as they are mostly influenced by the assumed particle Mie scattering. The detailed structures, including the brightening at high phase angles, are controlled by cloud particle size as it influences the single scattering phase functions and the interplay with the variation in observation geometry with orbital phase \citep[c.f.][]{Seager2000}. Thus a measurement of the orbital phase curve would provide constraints on cloud structure and composition, although obtaining the necessary precision for such observations would be challenging.

The spectral window probed by \jwst{}/NIRISS-SOSS is very similar to \tess{} and in that it probes the redder wavelengths where the planets are darker. \plato{} will struggle similarly as the observational bandpass is only catching the very edge of the rise in albedo (see \autoref{fig:albedo}). \cheops{} has the bluest reaching bandpass of those discussed here and it may be possible to use \cheops{} in concert with \tess{} to disentangle thermal emission from reflect light assuming the requisite sensitivity can be reached. For a 9th magnitude G5 star the initial science requirements called for 10~ppm precision in 6 hours \citep{Broeg2013}. Observing with \cheops{} has the same trend as the strawwoman blue filter, i.e. increasing contrast with increasing $f_{\rm sed}$, but wavelength range limits expected contrast to no larger than 40~ppm for the cloudless planet M. \kepler{} would have performed similarly, except without the reddest wavelengths, peaking at 50~ppm for planet A. \ariel{} will also track similarly.

\added{As explained in \autoref{sec:models}}, the albedo spectra and phase curves computed here all assumed an atmospheric structure in equilibrium with the cloud decks expected from rainout-equilibrium chemistry (primarily ZnS, KCl, and $\rm Na_2S$). These are dark clouds with lower scattering albedos compared to, for example, water clouds and thus tend to darken the planets at wavelengths measured by \tess{} while water clouds have the opposite trend with sedimentation efficiency \citep{Batalha2018Color, Macdonald2018} and can rapidly brighten a planet. Additionally, the atmospheres considered here are good candidates in which to form sulfur-bearing hazes such as $\rm S_8$ \citep{Zahnle2016} and are expected to to have a dramatic impact on the albedo spectrum \citep{Gao2017, Batalha2018Color}. Finally, disequilibrium processes, notably photochemistry, can be expected to produce additional aerosol \added{and gaseous} species not considered here which can impact the computed spectra and phase curves.

\subsection{Thermal Contamination}

\begin{figure}
    \plotone{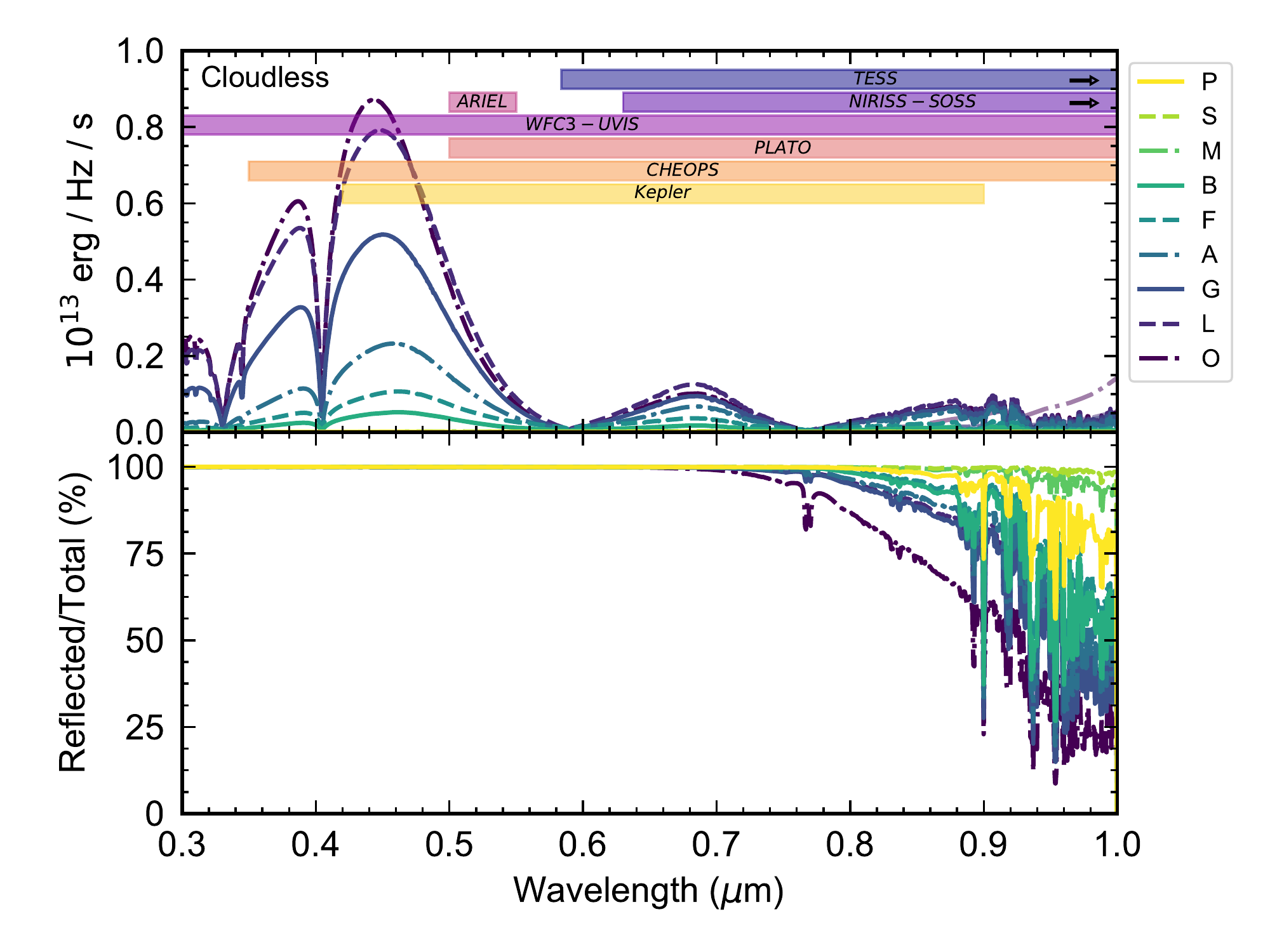}
    \caption{The predicted planetary reflected signal vs thermal emission. Top: the reflected light spectrum as measured in flux units. Note: We also plot emitted light but only planet O is visible in the lower right corner. Bottom: the percent contribution of reflected light to the total flux as a function of wavelength. We assume the star emits as a blackbody. \label{fig:ratio}}
\end{figure}

While the contrast provides an idea of the detectability of a planet relative to its host star (particularly in direct imaging scenarios) it is important to remember that stars hosting the Nine are cool stars. Sensitivity becomes a larger issue for follow up campaigns of transiting planets and redder bandpasses optimize stellar signal to noise ratios. We can compute the percentage of the total flux that comes from the reflected light of the planet as a function of wavelength versus the emitted light from the planet as computed by the structure model. The result for the cloudless case is shown in \autoref{fig:ratio}. For the hottest planets in the sample, thermal contamination can become significant at wavelengths longer than 0.8~$\mu$m.

\begin{figure}
    \plotone{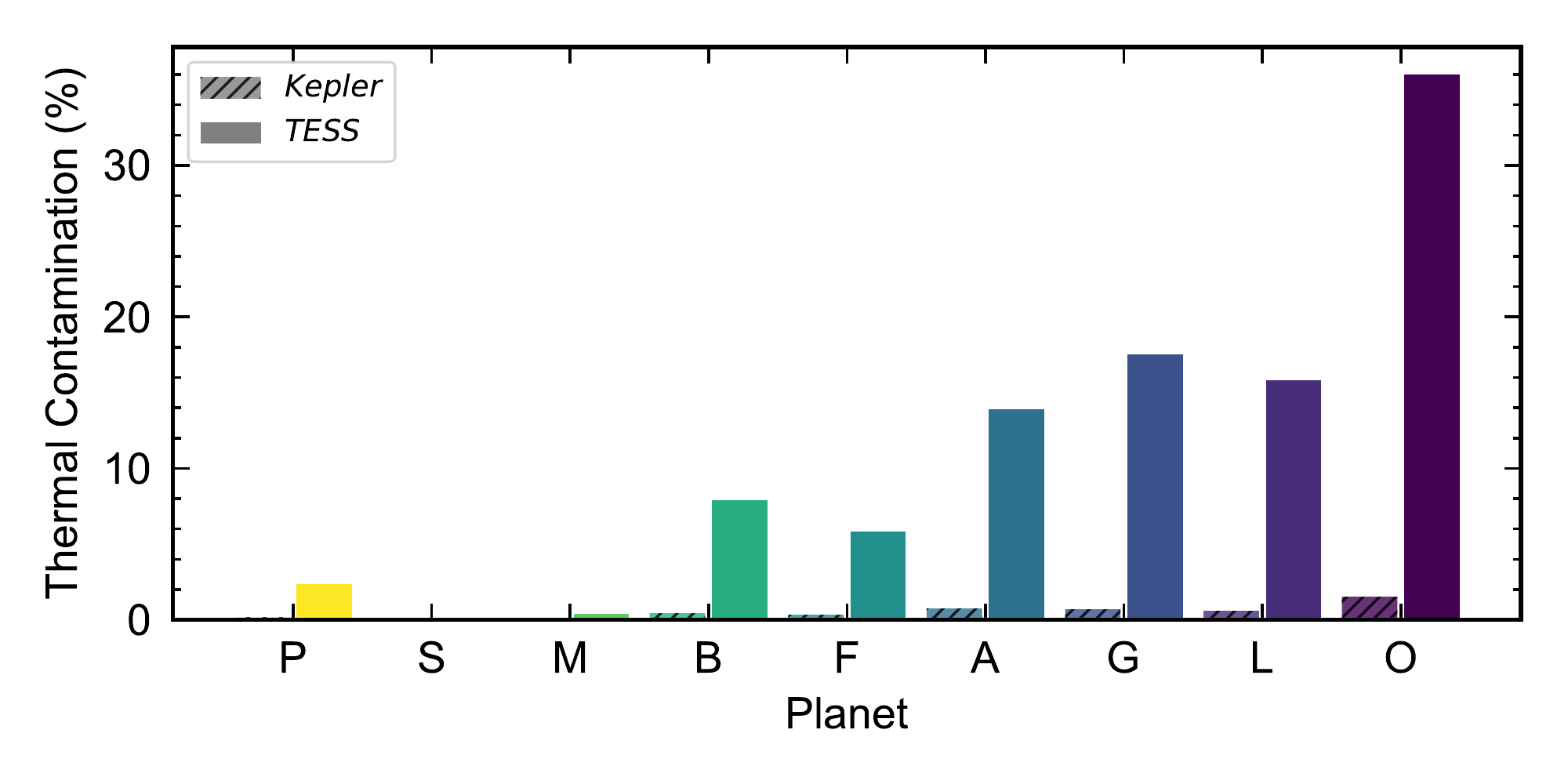}
    \caption{The predicted thermal emission contamination in percent assuming a cloudless atmosphere. The hashed bars represent the contamination \kepler{} would have seen and the solid bars represent the contamination \tess{} would see.\label{fig:thermperc}}
\end{figure}

We can consider the contamination in a particular bandpass by filter-integrating the reflection and emission spectra for each planet. The results are shown in \autoref{fig:thermperc} for \kepler{} and \tess{}. Recall, that we selected our planets so that $F_{\rm R}/F_{\rm E} >$  10 assuming a geometric albedo of $A_{\rm g}$=0.3 at all wavelengths. This would have limited the thermal contamination to less than $\sim$9\%. We estimate that \kepler{} would have measured a reflected light signal with less than 2\% thermal contamination and \cheops{} would measure at most 4\%.

The cloudless versions of the Nine are less reflective and thus the hottest planets have more thermal contamination than we estimated. This is caused by flux emerging from deeper and hotter layers where the temperature is greater than the equilibrium temperature. The addition of clouds even at the highest sedimentation efficiencies greatly reduce the thermal contamination to far below the percent level by blocking this opacity window. \added{Relaxing this assumption in our initial selection criteria could have added additional planets with higher insolations.}

\section{Conclusions}
\label{sec:conc}
We modeled the atmospheres of nine planets and generated albedo spectra with varying cloud sedimentation efficiencies. These planets\added{, the Nine,} are archetypes for the \tess{} discovered planets that we first estimate and then show will have little to no thermal contamination in the \tess{} bandpass ($<1$\% when clouds are present). We compare and contrast the reflected light signal in the \tess{} bandpass with that of a bandpass more focused in the blue and find that contrast ratios are more favorable in a bluer bandpass ($C\sim$10--59~ppm and as high as 150~ppm) than in the \tess{} bandpass ($C<$10~ppm except in the most lofted cloud scenario where they may be brighter).

Our results indicate that planet contrast is sensitive to cloud properties and scales with sedimentation efficiency. Planets with low sedimentation efficiencies will be comparably dark in the blue as in the \tess{} bandpass. With increasing sedimentation efficiency the planets will get darker in \tess{} but brighter in the blue thus leaving a clear imprint of clouds in the planet's atmosphere.

Generally speaking the clouds present for the classes of planets examined here tend to brighten the planets at red optical wavelengths and darken the planets at bluer wavelengths. This is because the $\rm Na_2S$ clouds scatter more than a clear gas in the red, but do absorb somewhat in the blue. \replaced{Thus unlike the case in the solar system, high cloud decks do not always lead to bright planets and it is important to consider the optical properties of the condensing cloud species when estimating contrast for exoplanets.}{Thus unlike the typical case in the solar system, high cloud decks do not always lead to bright planets at optical wavelengths. The expected contrast of cloudy exoplanets consequently ultimately depends on the optical properties of the aerosol species. Those used here are described in detail in \citet{Morley2012}.} 

While it is possible to estimate which \tess{} planets may have a detectable phase curve signal in the \tess{} bandpass (in our sample the Neptunes with low insolation and small semi-major axes, planet S and M), there is no way of indicating \textit{a priori} what the reflected light signal in the blue will be. The largest reflected light signal in the blue came from the Jupiter with the smallest semi-major axis (planet A), edging out the low insolation Neptune with the smallest semi-major axis (planet M). The complementary bluer reflected light measurements are critical for determining the energy budget of the planet. Phase curves from upcoming missions, such as \cheops{} and \plato{}, and completed missions, like \emph{CoRoT} and \kepler{}, with bluer bandpass edges can potentially be used in concert with \tess{} phase curves to disentangle reflected light from emitted light as in \citet{Placek2016} in cases with more thermal contamination. For those with little to no contamination like the representatives in this work we can attain a more complete picture of the planet's atmosphere and the clouds present therein.

For planet's around cool stars, detecting reflected light is further complicated by the brightness and activity of the star at these wavelengths. Thus it \added{is} important to characterize the host stars and the variable high-energy environment they create. These representative planets are around cooler stars that would potentially lead to lower signal-to-noise observations if observed in the blue vs in the red but at the expense of planet contrast. \added{Observatories with redder bandpasses will need to be very precise to measure the few to tens of ppm reflected light signals here.}

\added{Of the bandpasses that we include here,} \hst{}/WFC3-UVIS is currently the only space-based observatory with the capability of measuring reflected light from planets like \replaced{these}{the Nine} without risk of thermal contamination. \added{Based on modeling work here,} the 0.3--0.5 window is critical because that is where the planets are most reflective. Future near-UV and optical instrumentation will be needed not only to measure reflected light but also to continue to study the effects of stellar activity on exoplanet atmospheres. While focused missions such as \tess{} are unable to make these measurements alone, future flagships with direct imaging capabilities should carefully consider their ability to meet the goals of characterization of exoplanet atmospheres in the near-UV and short optical wavelengths \added{since they will be able to make many phase curve observations}. 

\acknowledgements

The authors thank Professor Caroline Morley for unwanted insight that made N.~E.~B. do a bunch more work. We also thank Daniel Thorngren for providing us with the inflation limits to eliminate phantom inflated planets from the sample. Additionally, we thank Peter Gao for humorous comments that greatly improved this manuscript. This work could not have been completed without the fundamental opacity work by Richard Freedman and Roxana Lupu. Work performed by L.~C.~M. was supported by the Harvard Future Faculty Leaders Postdoctoral fellowship. N.~E.~B. acknowledges support from the University of California President's Postdoctoral Fellowship Program. This research has made use of the NASA Exoplanet Archive, which is operated by the California Institute of Technology, under contract with the National Aeronautics and Space Administration under the Exoplanet Exploration Program.

\added{\facility {Exoplanet Archive}}

\software{Astropy \citep{Astropy}, PandExo \citep{Batalha2017b}, PySynphot \citep{PySynphot}, NumPy \citep{NumPy}, Pandas \citep{Pandas}, Scikit-Learn \citep{sklearn}, Matplotlib \citep{matplotlib}}

\bibliographystyle{aasjournal}
{\tiny \bibliography{Mendeley}}
\end{document}